\documentclass[english,a4paper,leqno,11pt]{article}

\usepackage{graphicx}

\begin{document}
\title{Monotonically convergent optimal control theory of quantum systems with spectral constraints on the control field}
\author{M. Lapert \thanks{Institut Carnot de Bourgogne, Dijon, France}
\and R. Tehini
\thanks{Institut Carnot de Bourgogne, Dijon, France}
\and G. Turinici
\thanks{CEREMADE, Universit\'e Paris Dauphine, Place du
Mar\'echal De Lattre De Tassigny, 75775 Paris Cedex 16, France}
\and D. Sugny \thanks{Institut Carnot de Bourgogne, Dijon, France}}
%\email{dominique.sugny@u-bourgogne.fr}
\date{\today}
\maketitle
\begin{abstract}
We propose a new monotonically convergent algorithm which can enforce
spectral constraints on the control field (and extends to arbitrary filters). The
procedure differs from standard algorithms in that at
each iteration the control field is taken as a
linear combination of the control field (computed by the standard algorithm)
and the filtered field.
The parameter of the linear combination is chosen to
respect the monotonic behavior of the algorithm and to be as close
  to the filtered field as possible.
We test the efficiency of this method on molecular alignment.
Using band-pass filters, we show how to select particular
rotational transitions to reach high alignment efficiency. We also
consider spectral constraints corresponding to experimental
conditions using pulse shaping techniques. We determine an optimal
solution that could be implemented experimentally with this
technique.
\end{abstract}
%\pacs{32.80.Qk,37.10.Vz,78.20.Bh}
\section{Introduction}
Quantum control is a field of growing interest both from the
experimental and theoretical points of views
\cite{warren,rabitz0,nielsen}. On the experimental side, the
development of pulse shaping techniques opens new ways to control
atomic or molecular processes by laser fields. Many promising
results have been obtained in the implementation of closed-loop
control experiments (CLC) based on genetic or evolutionary
algorithms (See \cite{baumert,daniel,judson,assion,levis} to cite
a few). This setup is made of a pulse shaper controlled by the
algorithm which from the results of the preceding experiments
builds an improved new control field. Such algorithms lead to very
efficient solutions, but have some negative points. In particular,
no insight into the control mechanism is gained from this approach
since no knowledge (or a minimal one) about the system is needed
and the control field is not optimal by construction.

On the theoretical
side, optimal control theory (OCT) is a powerful tool to design
electric fields to control quantum dynamics
\cite{shapiro,rice,tannorbook}. Numerical aspects of optimal
control theory have been largely explored. For simple systems with
few energy levels, geometric aspects of optimal control based
on the Pontryagin maximum principle \cite{jurdjevic,bonnard} have
also been investigated \cite{boscain1,sugnynew,mesure}.
Monotonically convergent algorithms is other efficient approach
to solve the optimality equations
\cite{tannor,rabitz2,zhu,zhu0,maday,ohtsuki1,ohtsuki2,sugny2,bifurcating,sugny7}.
They have been applied with success to a large number of
controlled quantum systems in atomic or molecular physics
 and in quantum computing. These methods are flexible and can be adapted to different non
standard situations encountered in the control of molecular
processes. Among recent developments, we can cite the question of
 nonlinear interaction with the control field
\cite{lapert,nakagami,nakagami2,tehini} and the question of
spectral constraints on the field
\cite{gollub,werschnik,artamonov,hornung,gross}. This latter
problem is particularly important in view of experimental
applications since not every control field can be
produced by pulse shaping techniques
\cite{baumert,daniel,judson,assion,levis}. For instance, liquid
crystal pulse shapers are able to tailor only a piecewise constant
Fourier transform of the control field in phase and in amplitude.
Experimentally, the spectral amplitude and phase are discretized,
e.g. into 640 points which is the number of pixels in a currently
used standard mask. In this context, a challenging question is to incorporate advantages and efficiency of OCT into CLC. OCT should provide insights into the control mechanisms and speed up the convergence of the experimental algorithm towards an optimal solution. Note that the theoretical analysis of the control dynamics is more efficient if the computational schemes can include experimental constraints directly in the algorithm. A brute force strategy which consists in applying the constraints after the optimization leads to worse results.

This paper aims at taking a step towards the answer to this question. For that purpose, we present a monotonically convergent algorithm
which can take into account spectral constraints on the control
field. Unlike other proposals in the literature~\cite{gollub,werschnik,artamonov,hornung,gross}
our approach exactly enforces the monotonicity property which is important to ensure the convergence of the algorithm. The construction of a monotonic
algorithm with spectral constraints is a very difficult task, since  constraints in the frequency domain
require nonlocal knowledge of the control field at
 all times whereas this field is computed progressively in time by the algorithm.
These two requirements are incompatible. This difficulty has been
bypassed in~\cite{gollub} for finite impulse response filters by using the control field of the
preceding iteration to construct the Fourier transform in order to
satisfy spectral constraints. Here we propose a similar
algorithm but with a modified electric field. This means that we
consider a standard monotonically convergent algorithm
\cite{zhu,zhu0,maday} giving at each iteration a control field. In
particular, only the energy of the field is penalized by the cost
which does not depend on spectral constraints. We then apply a filter to this
field to obtain a filtered  control field. We next construct
a new field as a linear combination of these two fields.
The parameter of the linear combination is such that the algorithm
remains strictly monotonic and the new control field is as close
to the filtered field as possible. As expected, the filtering
comes with the price of a slow down in convergence.

We test the efficiency of this approach on molecular alignment
which is a well-established topic in quantum control \cite{friedrich,seideman,stapelfeldt}. Molecular alignment has been optimized both experimentally \cite{exp1,exp2,exp3,exp4} and theoretically in different studies using genetic algorithms \cite{shir,hertz,rouzee} or optimal control theory
\cite{salomon,pelzer}. Note that previous optimal control studies have not
considered spectral constraints. The use of spectral constraints will allow one
to construct new optimal control fields to reach a high alignment efficiency. In \cite{salomon},
it was shown that aligned states are reached by rotational ladder climbing, i.e., by successive rotational excitations.
Using spectral constraints, we determine an optimal control field that induces only particular rotational transitions. This shows how to guide the
algorithm towards a particular mechanism. In a second step, we consider spectral constraints corresponding to experimental pulse shaping techniques. Our
new algorithm leads to an optimal solution that could be implemented experimentally. The solutions determined from genetic algorithms in \cite{shir,hertz,rouzee}
do not correspond to an optimal solution both maximizing the alignment and minimizing the energy of the control field. Minimizing the intensity of the field
 can be interesting to avoid parasitic phenomena such as ionization. Finally, we recall that in molecular alignment, due to the rapid oscillations of
the laser field, the effect of the permanent dipole moment
averages to zero and plays no role in the control of the dynamics.
The interaction between the molecule and the electric field is
therefore of order 2. Since the interaction is nonlinear, we use
the monotonic algorithm introduced in \cite{lapert} with a
non-standard cost of power 4 in the electric field to compute the
optimal solution.

The paper is organized as follows. In Sec. \ref{sec2}, we present
the problem of laser induced alignment of a linear molecule by
non-resonant laser fields. We describe how monotonically
convergent algorithms can be applied to such systems. We explain
how to modify the algorithm to take into account spectral
constraints. We show the efficiency of this new approach in Sec.
\ref{sec3} for a filtering which corresponds either to three particular rotational
frequencies or to experimental conditions using pulse shaping
techniques. Some conclusions and discussions are presented in Sec. \ref{sec4}. Some technical computations are reported in
appendix \ref{optapp}.
\section{Optimal control of molecular alignment with spectral
constraints}\label{sec2}
\subsection{Description of the model}
We consider the control of a linear molecule by a non-resonant
linearly polarized laser field of the form
$\vec{\mathcal{E}}(t)=\vec{E}(t)\cos(\omega t)$ where $\omega$ is
the carrier wave frequency and $\vec{E}(t)$ the laser pulse
envelope \cite{friedrich,seideman,stapelfeldt}. In the case of a
zero rotational temperature, the dynamics of the system is
governed by the Schr\"odinger equation. In a high-frequency
approximation \cite{friedrich}, this equation can be written as
\begin{equation}\label{eq1}
i\frac{\partial}{\partial
t}|\psi(t)\rangle=[BJ^2-\frac{1}{4}E(t)^2(\Delta
\alpha\cos^2\theta+\alpha_\perp)]|\psi(t)\rangle
\end{equation}
where $B$ is the rotational constant and $\Delta
\alpha=\alpha_\parallel-\alpha_\perp$ is the difference between
the parallel and perpendicular components of the polarizability
tensor. We use atomic units unless otherwise
specified. For numerical applications we consider the molecule
$CO$ with the following parameters $B=1.931$ $cm^{-1}$,
$\alpha_\parallel=15.65$ and $\alpha_\perp=11.73$ in atomic units
\cite{sekino}.
\subsection{Monotonically convergent algorithm}
The optimal control problem is solved by a monotonically
convergent algorithm. The goal of the control is to maximize the
projection of the system at time $t_f$ onto a target state where $t_f$ is
the duration of the control. The target state is
taken as the state which maximizes the alignment in a reduced
finite-dimensional Hilbert space spanned by the $j_{opt}$ first
rotational levels \cite{sugny3,sugny4,sugny5}. The reduced Hilbert
space is denoted $\mathcal{H}_{j_{opt}}$. We assume in this paper that $j_{opt}=8$.
A basis of the Hilbert space is given
by the spherical harmonics $|j,m\rangle$ with $j\geq 0$ and $-j\leq m\leq j$.
In case of pure state systems at a temperature $T=0~\textrm{K}$, the target state
$|\phi_f\rangle$ is simply the eigenvector of maximum
eigenvalue of the projection of
the operator $\cos^2\theta$ onto $\mathcal{H}_{j_{opt}}$. The optimal
control of molecular orientation and alignment has been considered
in a series of articles \cite{salomon,pelzer,pelzer2,lapert}. The
novelty of our approach lies in the fact that we will consider
in addition spectral constraints which leads to new optimal control fields. In other words, this means that there exists no unique optimal control field to reach a given target state and that it is possible to select an optimal solution via spectral constraints. Due to the interaction of order
2 between the system and the field, we use the algorithm
introduced in \cite{lapert} with a cost which is quartic in the field.

We consider the initial state $|\phi_0\rangle=|0,0\rangle$ and the following cost functional:
\begin{equation}\label{eq2}
J=|\langle
\phi_f|\psi(t_f)\rangle|^2-\int_0^{t_f}\lambda(t)E(t)^4dt
\end{equation}
where $\lambda(t)=\lambda_0/s(t)$ is a positive function.
Following \cite{sundermann}, we will choose $s(t)=\sin^2(\pi
t/t_f)$ which penalizes the amplitude of the pulse
at the beginning and at the end of the control. Here $\lambda_0$ is
chosen to express the relative weight between the projection and the
energy of the electric field. The augmented cost functional
$\bar{J}$ is defined as follows
\begin{eqnarray}\label{eq2}
\bar{J}&=&|\langle
\phi_f|\psi(t_f)\rangle|^2-\int_0^{t_f}\lambda(t)E(t)^4dt
-2\textrm{Im}[
\langle\psi(t_f)|\phi_f\rangle\int_0^{t_f}\langle
\chi(t)|(i\frac{\partial}{\partial t}-H)|\psi(t)\rangle dt]
\end{eqnarray}
where $|\chi(t)\rangle$ is the adjoint state and $\textrm{Im}$
denotes the imaginary part. Setting the variations of $\bar{J}$
with respect to $|\psi(t)\rangle$, $|\chi(t)\rangle$ and $E(t)$
to be zero implies that $|\psi(t)\rangle$ and $|\chi(t)\rangle$ satisfy the
Schr\"odinger equation with respective initial conditions
$|\phi_0\rangle$ and $|\phi_f\rangle$:
\begin{eqnarray}
\begin{array}{ll}
(i\frac{\partial}{\partial t}-H(t))|\psi(t)\rangle =0 \nonumber\\
(i\frac{\partial}{\partial t}-H(t))|\chi(t)\rangle =0 \nonumber
\end{array} .
\end{eqnarray}
The optimal electric field is solution of the polynomial equation:
\begin{eqnarray}\label{eq3}
4\lambda(t)E(t)^3
+\textrm{Im}[\langle\psi(t_f)|\phi_f\rangle\langle\chi(t)|(\Delta
\alpha\cos^2\theta+\alpha_\perp)E(t)|\psi(t)\rangle]=0.
\end{eqnarray}

We solve this set of coupled equations by using a monotonic
algorithm. To simplify the presentation of the algorithm
\cite{lapert}, we consider that the forward and backward electric
fields are the same. The iteration is initiated by a trial field
$E_0(t)$. Let us assume that at step $k$ of the iterative
algorithm the system is described by the triplet
$(|\psi_k(t)\rangle,|\chi_{k-1}(t)\rangle,\tilde{E}_k(t))$
associated to the cost $J_k$ given by
\begin{equation}\label{eq4}
%J_k(\tilde{E}_k)
J_k = J(\tilde{E}_k)
=|\langle
\phi_f|\psi_k(t_f)\rangle|^2-\int_0^{t_f}\lambda(t)\tilde{E}_k(t)^4dt.
\end{equation}
The spectral constraints are defined through a filter
$\mathcal{F}$ in the frequency-domain. This means that for a given electric field
$E(t)$, $\mathcal{F}[E(t)]$ only contains the  admissible frequency components of the field.

We determine the triplet at step $(k+1)$ from the triplet at step $k$
by the following operations. We first propagate backward the adjoint state $|\chi_k(t)\rangle$ with the field $\tilde{E}_k(t)$ and initial condition $|\phi_f\rangle$:
\begin{equation}
i\frac{\partial}{\partial t}|\chi_k(t)\rangle
=H(\tilde{E}_k(t))|\chi_k(t)\rangle .\nonumber
\end{equation}
We then propagate forward the state $|\psi_{k+1}\rangle$ with
initial condition $|\phi_0\rangle$:
\begin{equation}
i\frac{\partial}{\partial t}|\psi_{k+1}(t)\rangle
=H(E_{k+1}(t))|\psi_{k+1}(t)\rangle ,\nonumber
\end{equation}
computing at the same time the electric field $E_{k+1}(t)$ by the procedure explained in the appendix \ref{optapp}.

Having computed the non-filtered optimal field $E_{k+1}(t)$, the next step of the algorithm consists in introducing the filtered field $\tilde{E}_{k+1}$ as:
\begin{equation}\label{eqGT1}
\tilde{E}_{k+1}(t) =
\tilde{E}_{k+1,\mu_{k+1}}(t)
\end{equation}
where we have denoted for any $\mu \in [0,1]$:
\begin{equation}\label{eq5}
\tilde{E}_{k+1,\mu}(t)=\mu
E_{k+1}(t)+(1-\mu)\mathcal{F}[E_{k+1}(t)]
\end{equation}
and $\mu_{k+1}$ is such
that the algorithm remains monotonic, i.e. such that
$\Delta J_{k+1}(\mu_{k+1})=J(\tilde{E}_{k+1,\mu_{k+1}}) -J(\tilde{E}_k)\geq 0$.
We choose a value of
$\mu_{k+1}$ such that $\Delta J_{k+1}(\mu_{k+1})>0$ but close to 0. We explain in
Sec. \ref{sec3} on the example of molecular alignment how to
determine numerically $\mu_{k+1}$. This choice ensures the monotonic
behavior of the algorithm and to be as close to the filtered field
as possible.
\section{Numerical results}\label{sec3}
We first present numerical results at $T=0~\textrm{K}$ to show the
efficiency of the algorithm for long control fields (Sec.
\ref{seclong}). Spectral constraints correspond in this case to
band-pass filters which allow to select particular rotational
transitions. We also analyze the temperature effects for shorter
control fields in Sec. \ref{nonzero} with constraints mimicking
the experimental pulse shaping techniques.
\subsection{Zero temperature}\label{seclong}
We consider a control field with a duration of ten rotational
periods. The $CO$ rotational period is denoted by $t_{per}$. This
long duration allows to simplify the structure of the Fourier
transform of the fields. The filter is composed of three bandpass
filters around the frequencies $4B$, $10B$ and $26B$ with a
bandwidth of $B/2$. Combining these three frequency components, we
obtain the different rotational transitions between the even
rotational levels which are populated during the control. We have
for instance that $\omega_{02}=10B-4B$, $\omega_{24}=10B+4B$ and
$\omega_{46}=26B-4B$ where $\omega_{JJ'}$ is the difference of
energy between the $J$th and the $J'$th rotational levels
\cite{nakagami2}. This shows that all the even rotational levels
can be populated using only these three frequency components.
Also, this means that the target state $|\phi_f\rangle$ can be
reached from the initial state $|\phi_0\rangle$. To compute the
parameter $\mu_{k+1}$ at each iteration, we have used a dichotomy
method to determine the zero, $\mu_0$, of $\Delta
J_{k+1}(\mu)=J(\tilde{E}_{k+1,\mu})-J(\tilde{E}_{k})$. The value
$\mu_{k+1}$ is the first value of $\mu$ given by this approach
such that $\Delta J_{k+1}(\mu)>0$ and $|\mu-\mu_0|\leq 0.01$. As
could be expected, the smaller the difference $|\mu-\mu_0|$ is,
the slower the convergence of the algorithm is.

\begin{figure}
\includegraphics[width=0.4\textwidth]{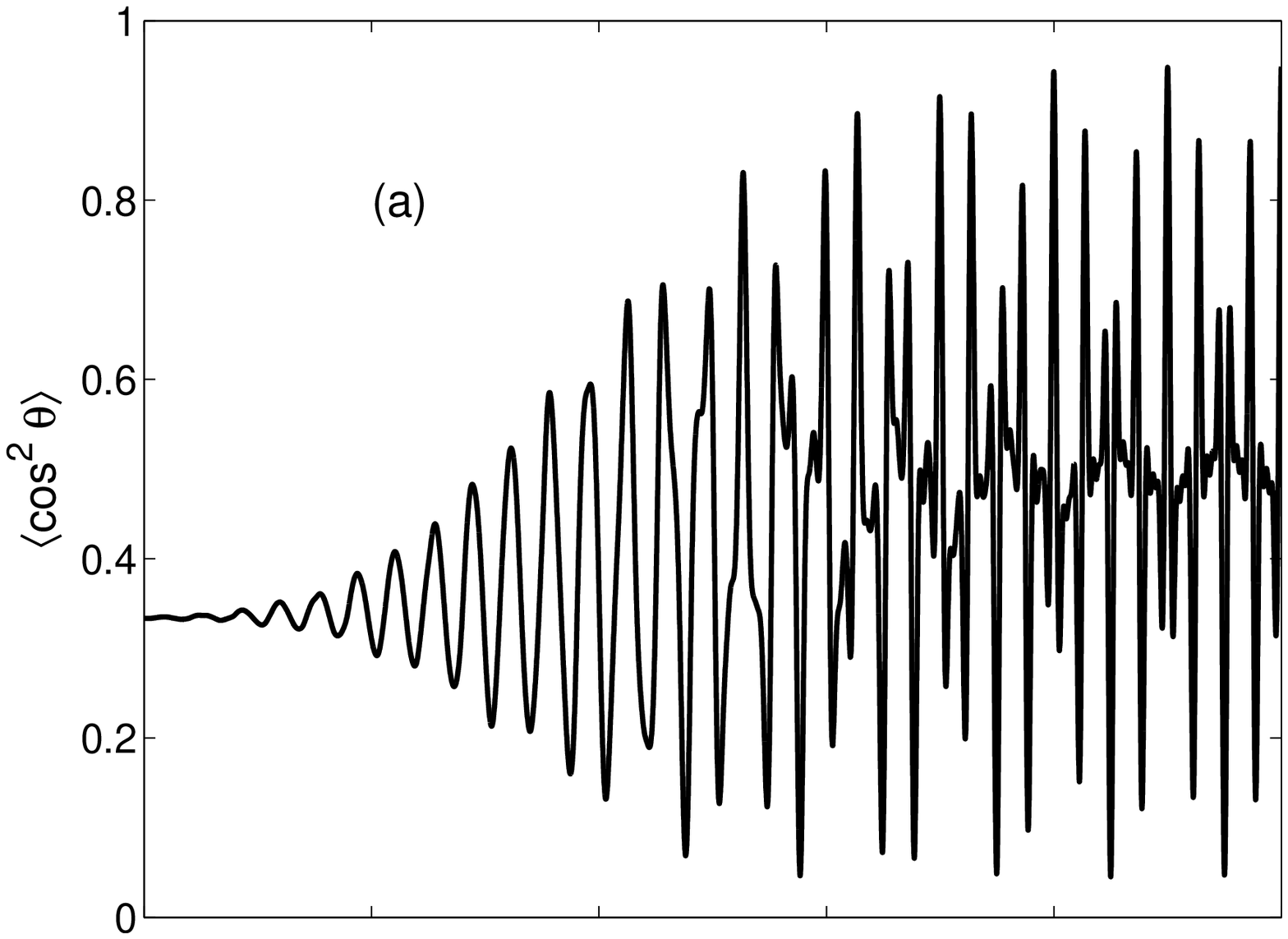}
\includegraphics[width=0.4\textwidth]{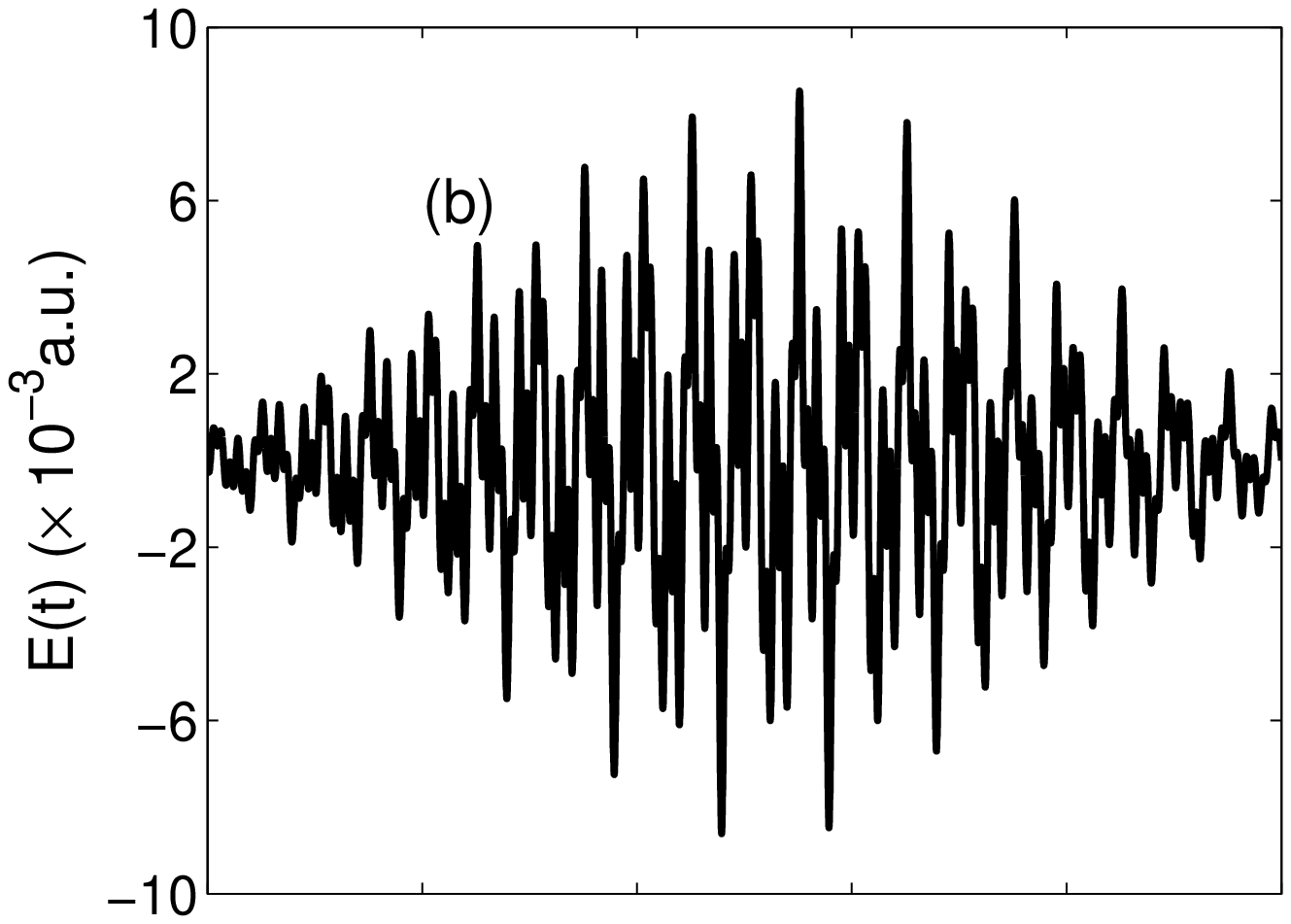}
\includegraphics[width=0.4\textwidth]{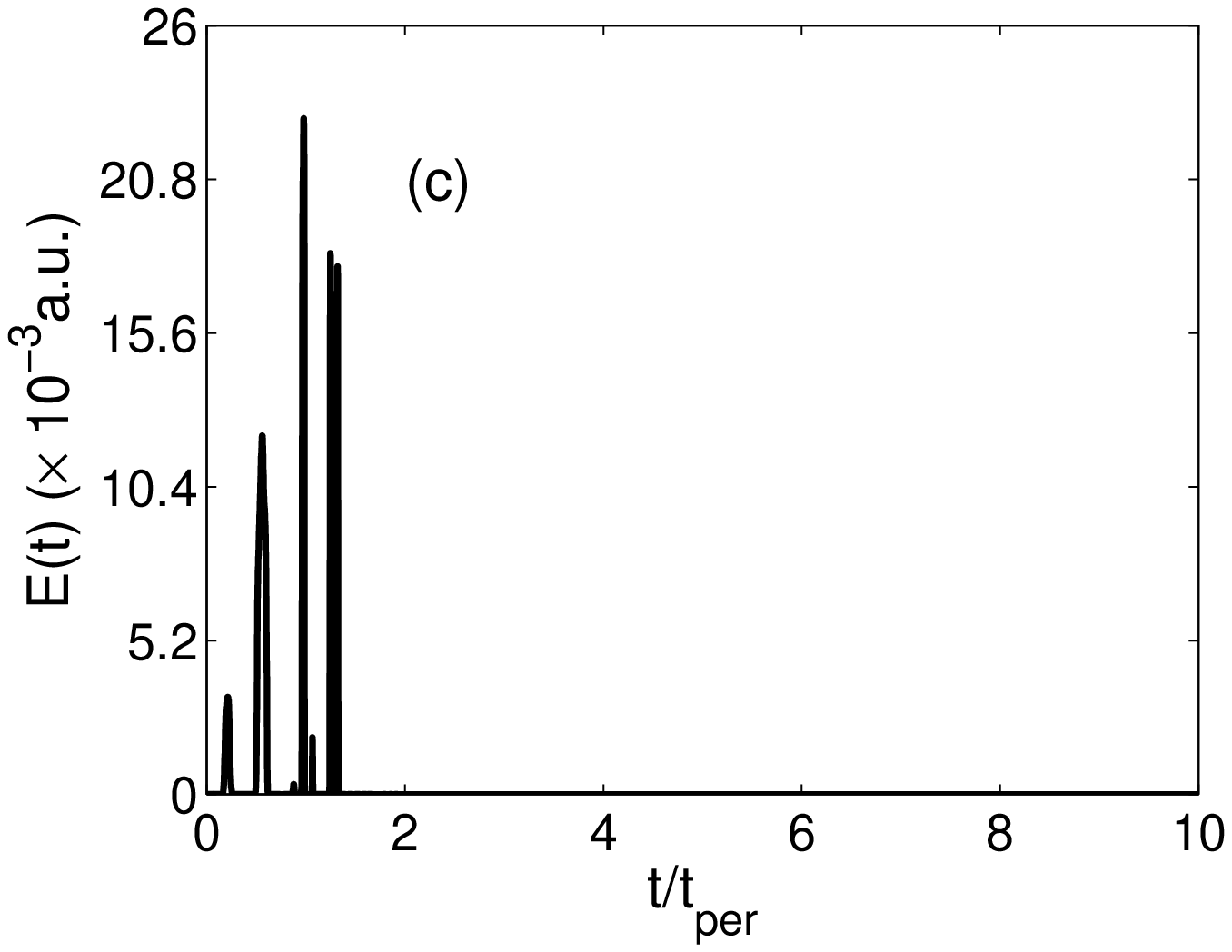}
\includegraphics[width=0.4\textwidth]{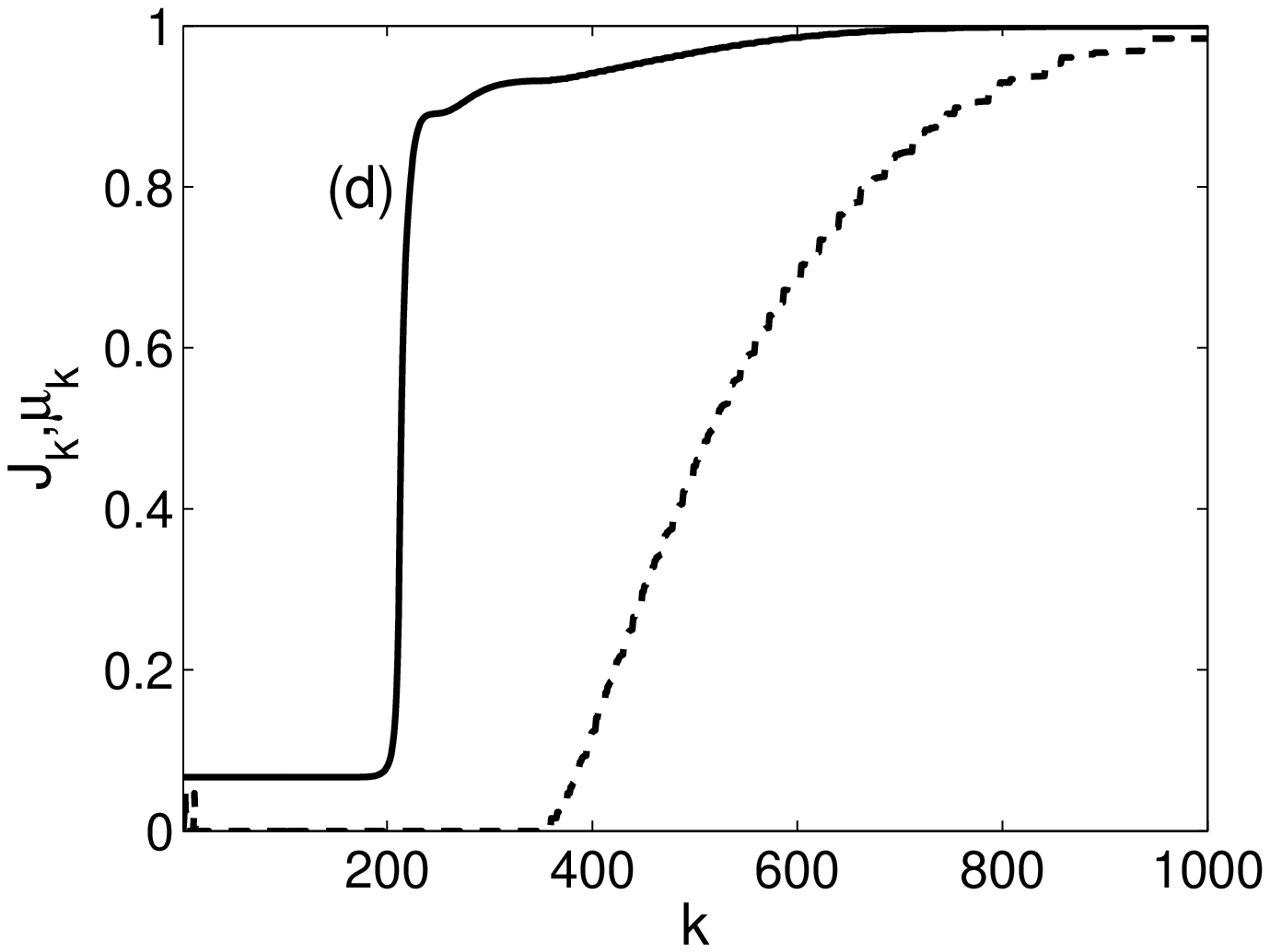}
\caption{\label{fig1} Plot as a function of the adimensional time
$t/t_{per}$ of (a) the expectation value
$\langle\cos^2\theta\rangle$ obtained from the algorithm with
filtering, (b) the corresponding optimal filtered field and (c)
the standard optimal field. (d) Plot as a function of the
number of iterations $k$ of the adimensional cost $J_k$ (solid
line) and of the parameter $\mu_k$ (dashed line) for the algorithm
with filtering.}
\end{figure}

\begin{figure}
\includegraphics[width=0.4\textwidth]{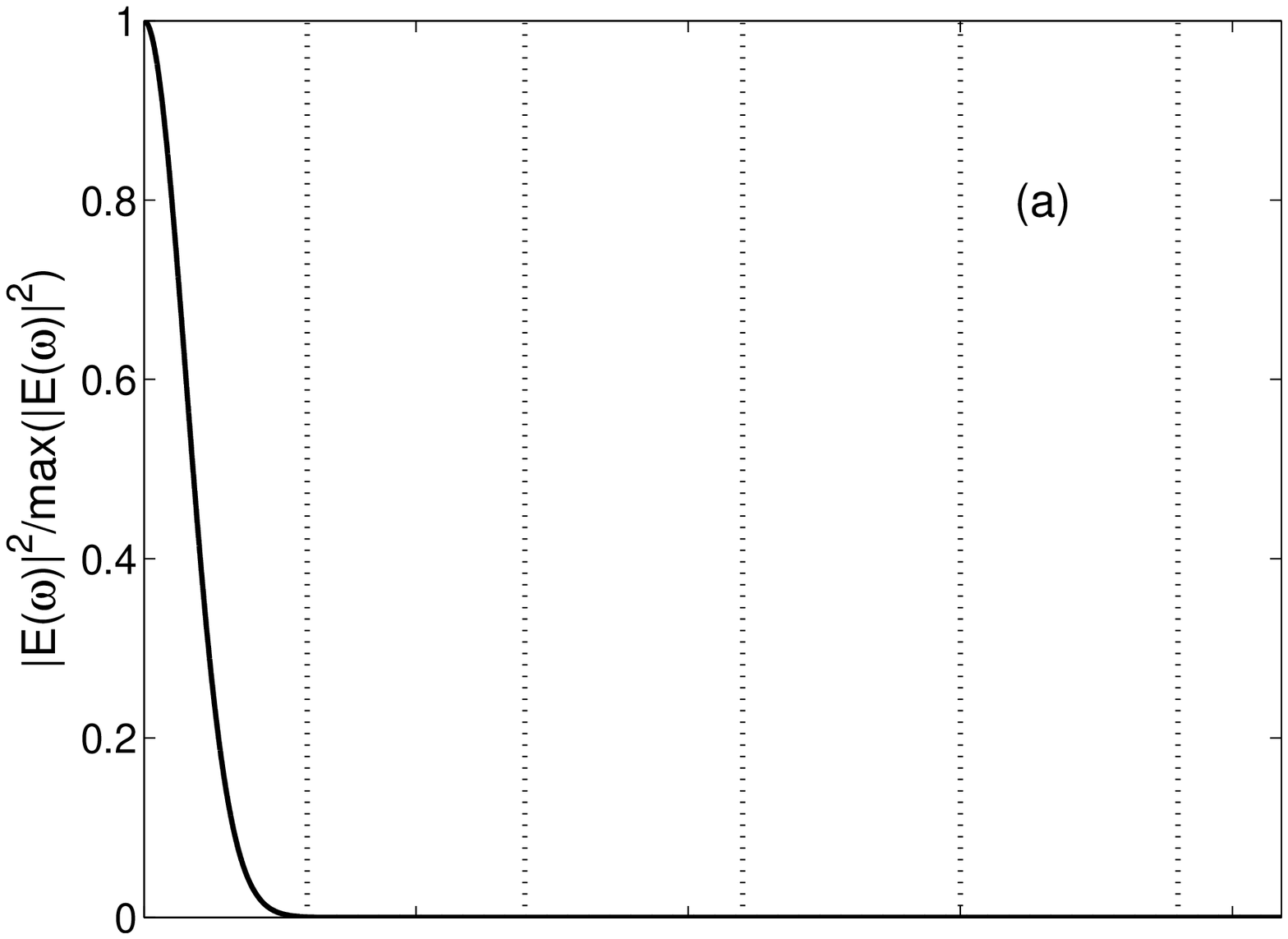}
\includegraphics[width=0.4\textwidth]{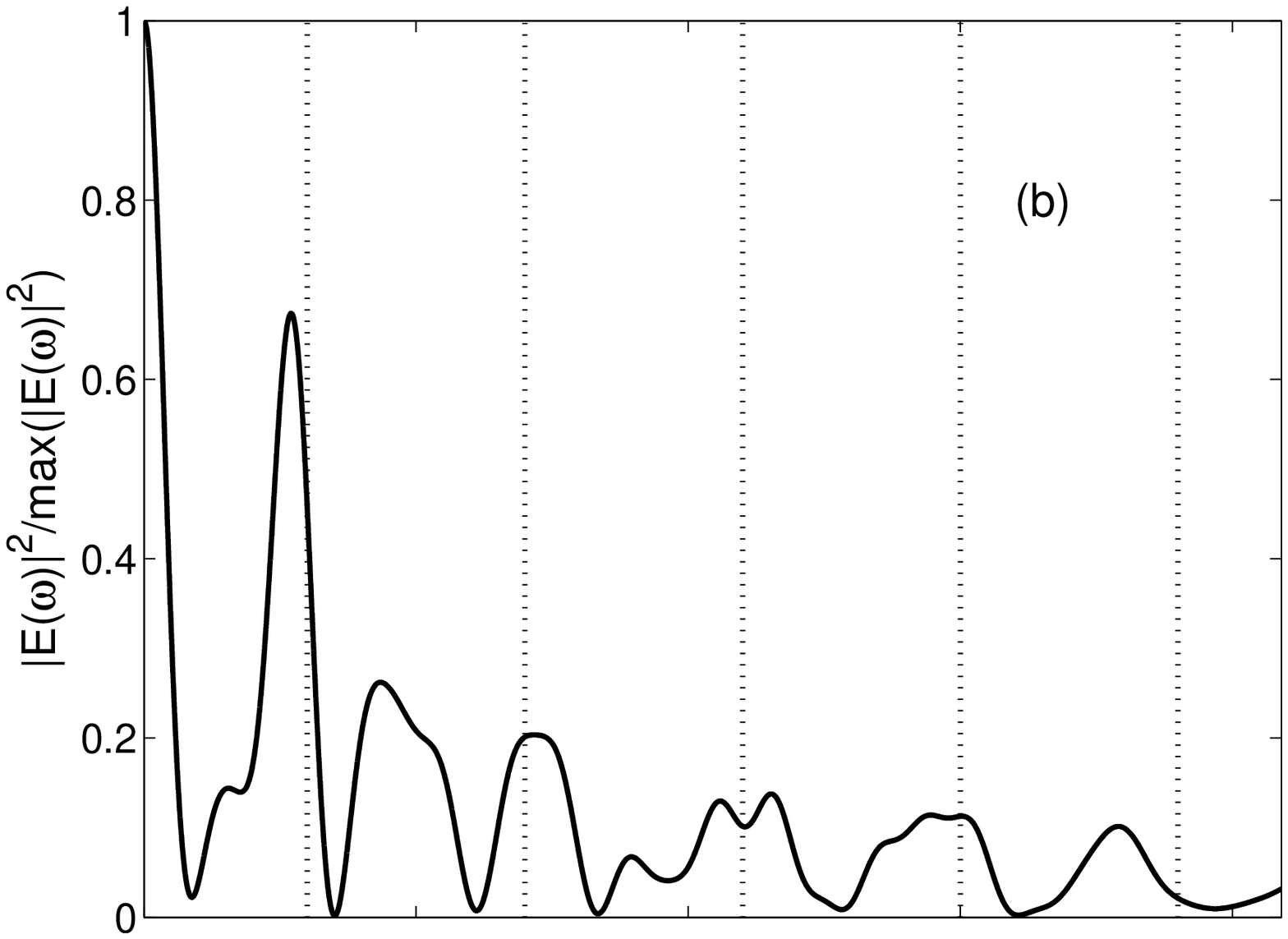}
\includegraphics[width=0.4\textwidth]{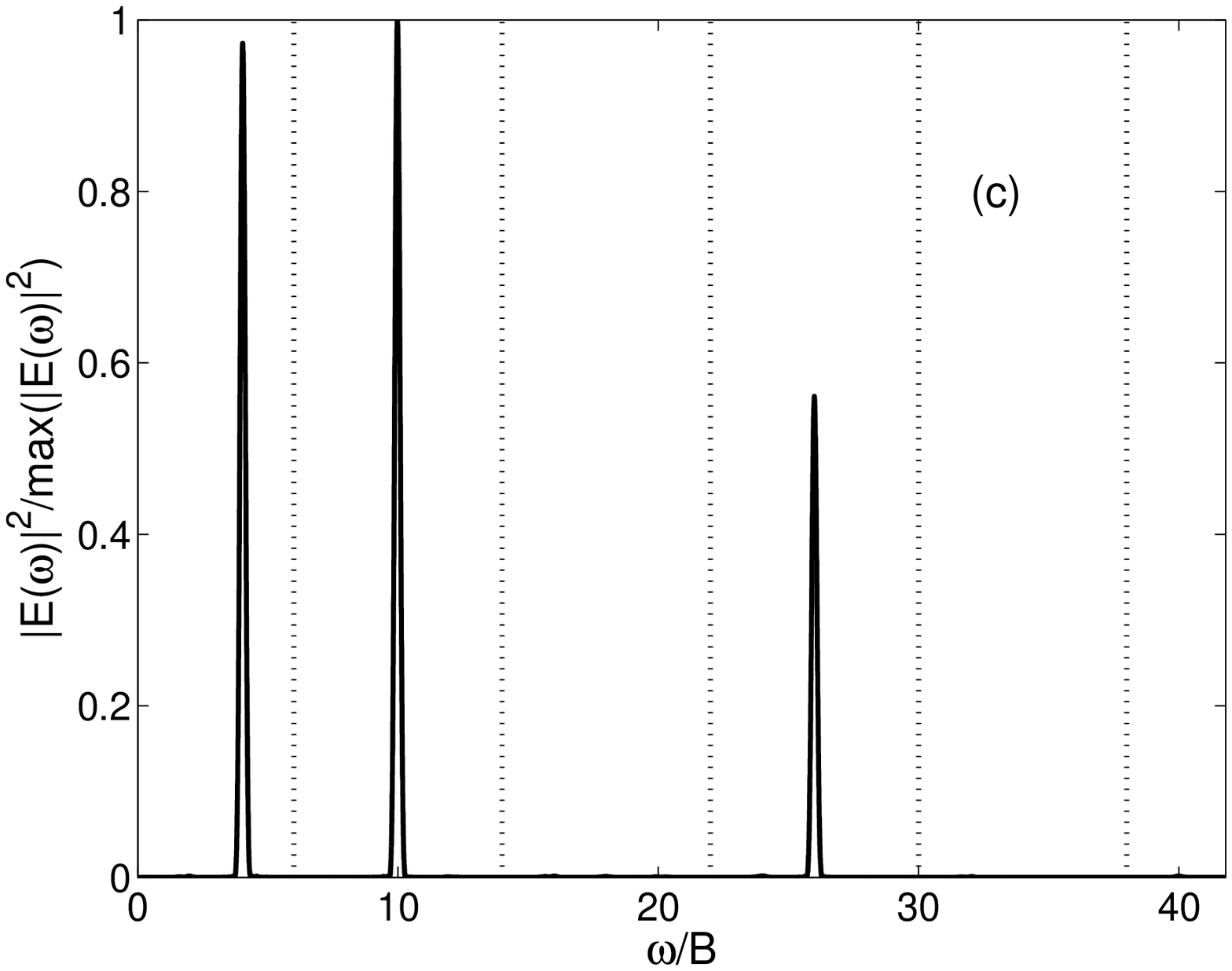}
\caption{\label{fig2} Normalized square modulus of the Fourier
transform of the trial field (a), of the optimal field obtained by
the standard algorithm (b) and of the optimal field obtained by
the algorithm with filtering. The vertical dashed lines correspond
to the rotational frequencies between two successive rotational
levels.}
\end{figure}

We apply the standard algorithm \cite{lapert} and the one with
filtering to maximize the projection onto the target state at time
$t=t_f$. The trial field is a Gaussian pulse with a duration
(FWHM) equal to 3 ps. The parameters of the algorithms are taken
to be $\lambda_0=1$ and $\eta=1$. The results of the computations,
i.e. the evolution of $\langle\cos^2\theta \rangle (t)$ and the
optimal control field, are presented in Fig. \ref{fig1} for the
algorithm with filtering. Very good results are obtained with a
final projection $|\langle \psi(t_f)|\phi_f\rangle |^2$ larger
than 0.99. Similar efficiencies have been reached by the standard
optimal field displayed in Fig. \ref{fig1}c. As could be expected,
the filtering slows down the convergence of the algorithm. We get
a projection close to 0.98 after 200 and 400 iterations for
respectively the standard and the modified algorithms. The
monotonic behavior of the new algorithm can be seen in Fig.
\ref{fig1}d. The evolution of the parameter $\mu_k$ as a function
of the number of iterations is also represented in Fig.
\ref{fig1}d. One sees that $\mu_k$ is equal to zero for the first
300 iterations and then it increases to reach a value close to 1
for $k\simeq 600$. Since after 300 iterations, the optimal field
for $\mu=1$ and the completely filtered one for $\mu=0$ are very
close to each other, the non-zero value of $\mu_k$ does not affect
the filtering of the algorithm. The discontinuous behavior of
$\mu_k$ in Fig. \ref{fig1}d is due to the dichotomy method and to
the criterion used to determine this parameter. Note that the
evolution of $\mu_k$ is the same in all the examples we have
considered, i.e. a zero value for the first iterations and then a
slow increase towards the value 1. Figures \ref{fig2} show the
Fourier transform of the optimal solutions obtained by the
standard algorithm and the one with filtering. The spectral
structure of the filtered optimal solution is very simple with
respect to the one without filtering. As expected, only three
frequency components appear in the Fourier transform of the
filtered field, whereas no frequency component can be clearly
distinguished for the optimal solution. One sees that the increase
of the parameter $\mu_k$ after 400 iterations does not affect the
Fourier structure of the optimal solution. Figure \ref{fig1}c
shows that the standard optimal solution does not use the ten
rotational periods since the control field vanishes after 1.75
periods. This short duration leads to a very complicated spectral
structure in Fig. \ref{fig2}b which contrasts with the very simple
structure of the filtered optimal field. The spectral constraints
can be viewed here as a tool for guiding the control dynamics
along a selected pathway in the frequency domain in the sense that
the standard optimal solution is a superposition of many pathways
leading to a complicated Fourier spectrum.
\subsection{Non-zero temperature}\label{nonzero}
In this section, we test the efficiency of the algorithm at a
non-zero temperature. The system is described by a density matrix
with a dynamics governed by the von Neumann equation
\cite{seideman,stapelfeldt}. The initial density matrix is the
Boltzmann equilibrium density operator at temperature $T$. We use
the target state $\rho_{opt}^{(j_{opt})}$ constructed in Ref.
\cite{sugny4} for a linearly polarized electric field. This
density matrix acts in the space $\mathcal{H}_{j_{opt}}$ with
$j_{opt}=8$ which is chosen with respect to the temperature and
the intensity of the field used. As before, we have used the
algorithm of \cite{lapert} with a cost functional similar to the
one of Sec. \ref{sec2}. We have considered three temperatures
$T=5,~7$ and 10 K and a shorter control field with a duration
equal to one rotational period. The numerical parameters of the
trial gaussian field are taken to be
$37.5~\textrm{TW}/\textrm{cm}^2$ for the peak intensity and 1 ps
for the duration (FWHM), except for $T=7~\textrm{K}$ where the
duration is equal to 0.475 ps. 6000 iterations have been used for
the first two temperatures and 8000 for the last one. The
parameters of the algorithm are equal to $\lambda_0=1$ and
$\eta=1$.

We have used a filtering which mimics experimental control
techniques using liquid crystal pulse shapers \cite{hertz}. Such
pulse shapers work in the frequency domain by tailoring the
spectral phase and amplitude of the electric field. This operation
can be defined through a filter $\mathcal{F}$ which transforms
over a given bandwidth the Fourier transform of the control field
into a piecewise constant function in phase and in amplitude. The
bandwidth is taken to be 7.28 THz which corresponds to two times
the 5th rotational frequency. A larger bandwidth has given
equivalent results. The filtered control field $\mathcal{F}(E(t))$
is finally obtained by an inverse Fourier transform. In the
following computations, the number of pixels $N$ is equal to 64,
128 or 256 both in amplitude and in phase. The pixels that
discretize the Fourier transform are taken equally spaced in the
frequency interval defined by the bandwidth.

To determine the parameter $\mu_{k+1}$, we compute at each
iteration $\Delta J_{k+1}(\mu)=J(E_{k+1,\mu})-J_k(\tilde{E}_k)$ for 10 values of
$\mu$ equally spaced in the interval $[0,1]$. Using these ten
values, we construct a polynomial fit of $\Delta J_{k+1}(\mu)$.
 We define $\mu_{k+1}$ as the smallest value of $\mu$ such that
$\Delta J_{k+1}(\mu_{k+1})=0.01\times \max_{\mu\in [0,1]}[\Delta J_{k+1}(\mu)]$.
We have checked that these ten values are sufficient.
\begin{figure}
\includegraphics[width=0.4\textwidth]{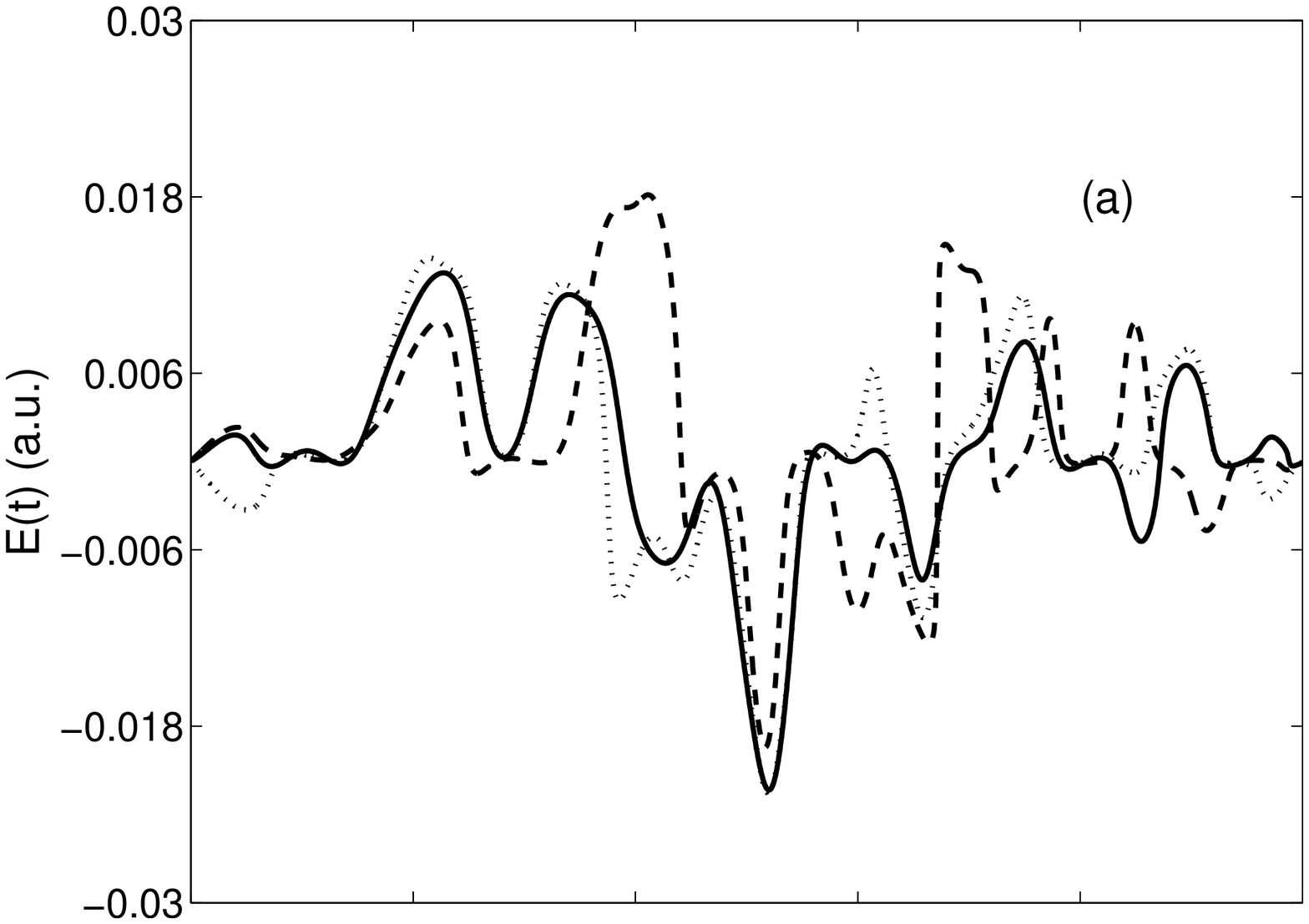}
\includegraphics[width=0.4\textwidth]{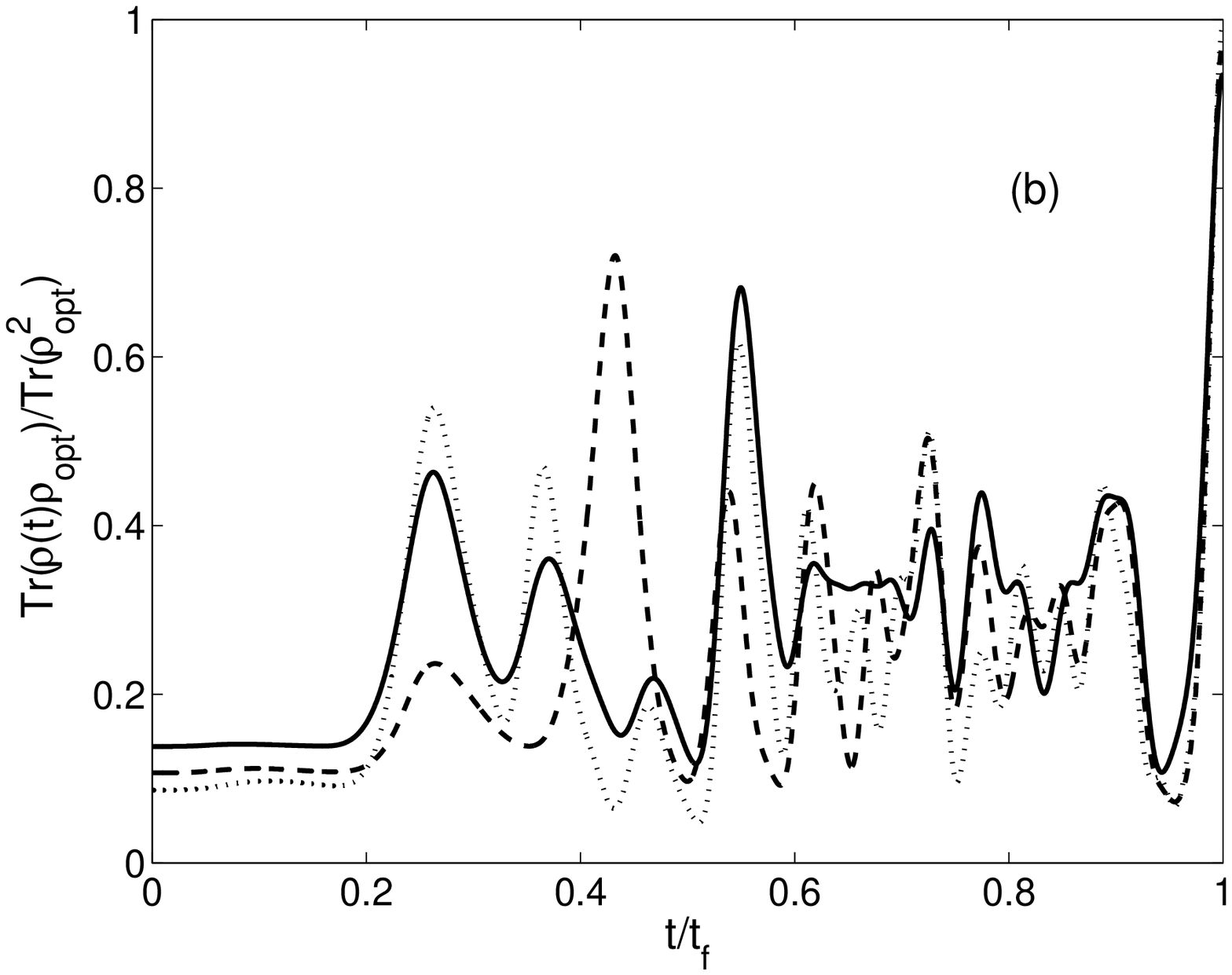}
\includegraphics[width=0.4\textwidth]{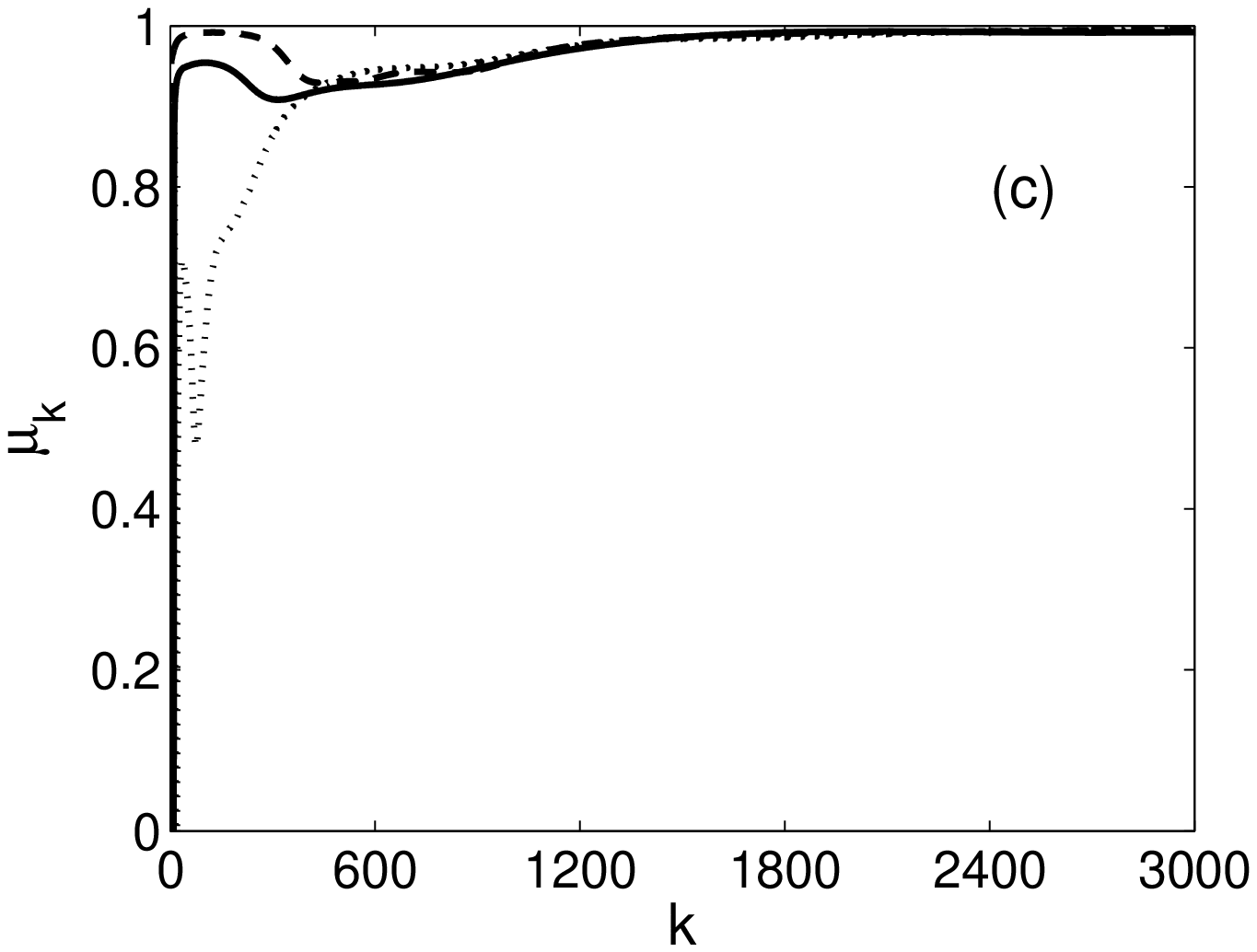}
\caption{\label{fig3} Plot as a function of the adimensional time
$t/t_f$ of (a) the optimal electric field and (b) the projection
onto the target state $\rho_{opt}$. (c) Plot as a function of the
number of iteration $k$ of the parameter $\mu_k$. Solid, dashed
and dotted lines correspond respectively to $T=10,~7$ and 5 K. The
number of pixels is equal to 128.}
\end{figure}
\begin{figure}
\includegraphics[width=0.4\textwidth]{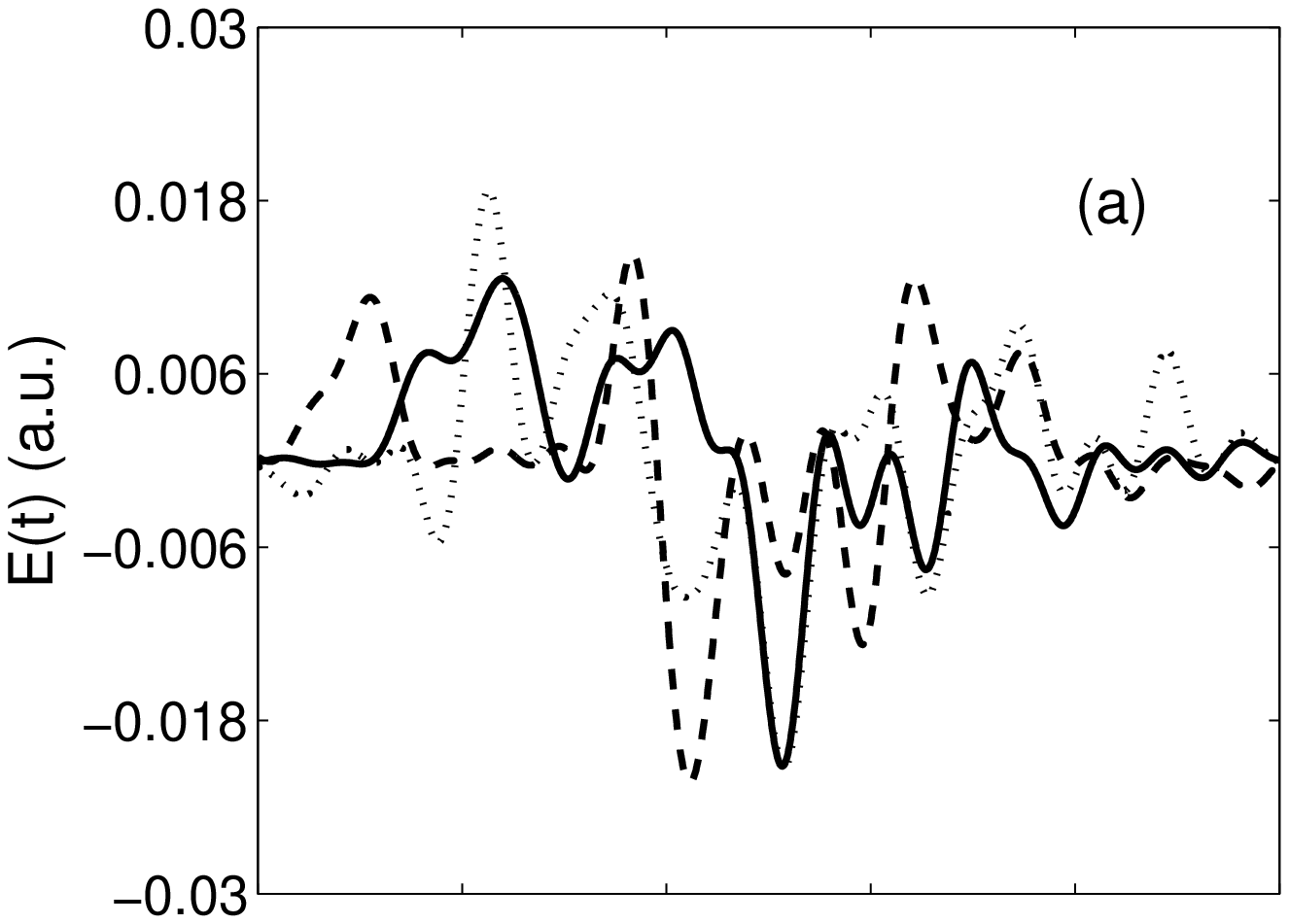}
\includegraphics[width=0.4\textwidth]{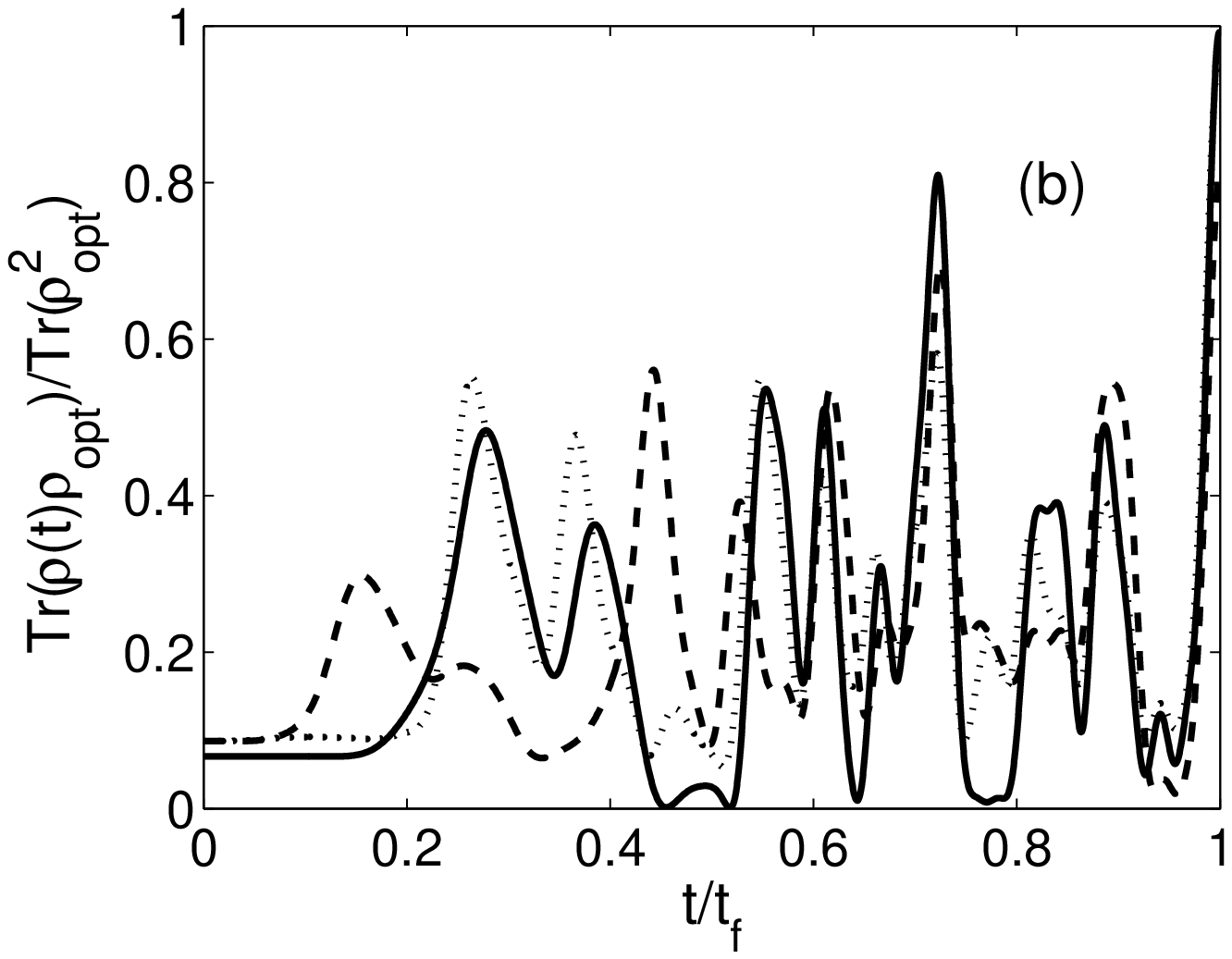}
\includegraphics[width=0.4\textwidth]{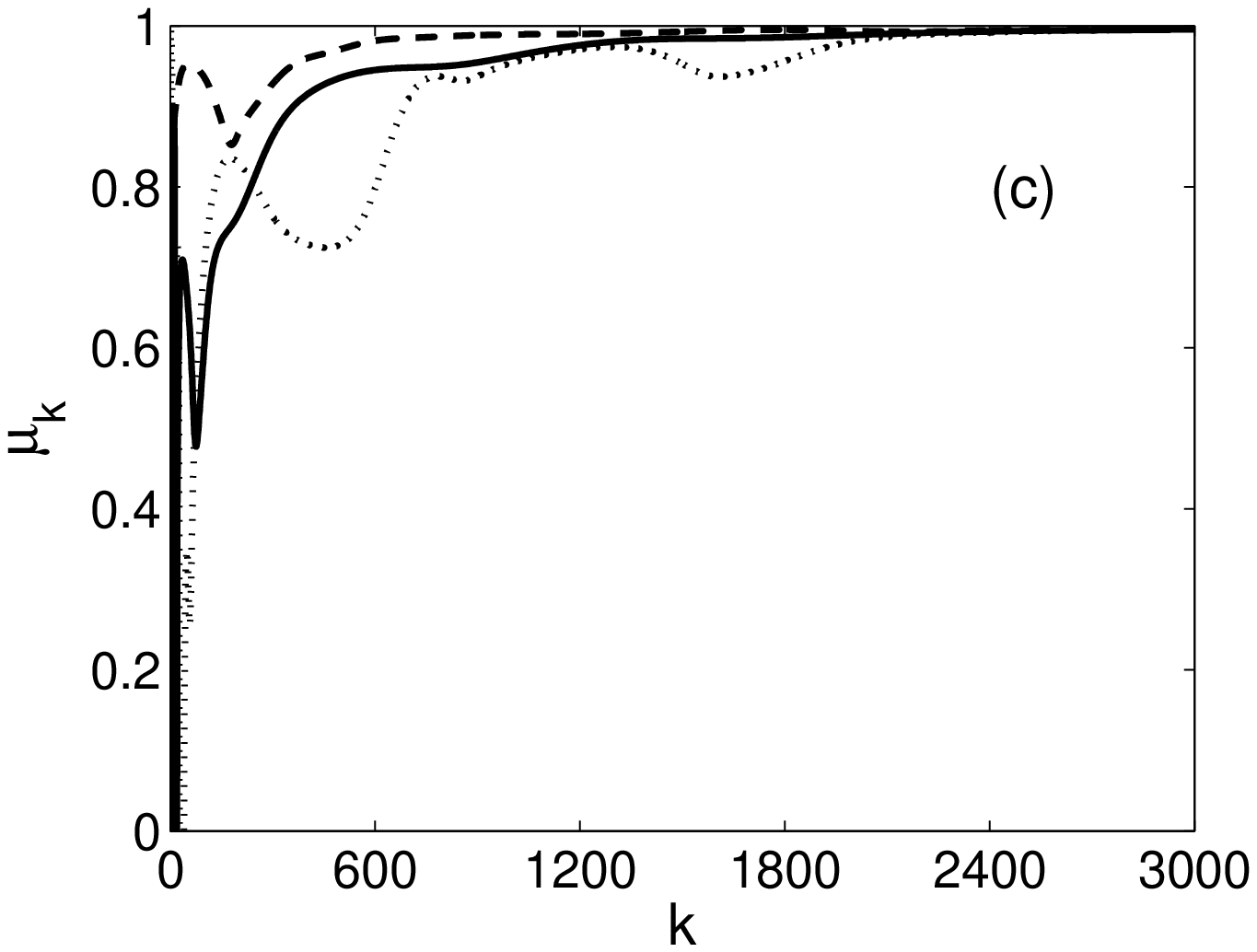}
\caption{\label{fig4} Same as Fig. \ref{fig3} but for $T=5$ K and
different numbers of pixels. Solid, dashed and dotted lines
correspond respectively to $N=128,~64$ and 256 pixels.}
\end{figure}
The results obtained for the three temperatures are displayed in
Fig. \ref{fig3} with a fixed number of pixels equal to 128. Very
efficient optimal fields have been constructed by the algorithm
since a final projection of 99.2\%, 97\% and 93.5\% has been
obtained for respectively $T=5,~7$ and 10 K. As can be expected,
the temperature has a negative effect on the alignment but the
control remains robust with respect to the temperature. Note also
the completely different structures of the three optimal fields
leading to three different evolutions for the projection. This
example shows that our algorithm can be used to construct
realistic control fields both in temperature and in intensity. In
Fig. \ref{fig3}c, one observes a similar evolution for the three
parameters $\mu_k$ which are close to zero for the first
hundred iterations and then quickly increase for $k\geq 100$. The
increase of $\mu_k$ is more pronounced for higher temperatures
which indicates that spectral constraints are more difficult to
impose when the temperature increases.

Figures \ref{fig4} illustrate the impact on the algorithm of the
number of pixels. We have obtained a final projection of
respectively 0.975, 0.989 and 0.992 for $N=64,~128$ and 256
pixels. In Fig. \ref{fig4}, we have represented the fields after a
final filtering to get solutions that can be implemented experimentally. We then
obtain 0.813, 0.892 and 0.926 for the three cases. The final result
passes for $N=64$ from 0.975 to 0.813 which shows that the optimal
solution does not satisfy very well the spectral constraints. One deduces
that $N=64$ is not a sufficient number of pixels to construct the
optimal field. This observation is also related to the rapid
increase of $\mu_k$ in Fig. \ref{fig4}c for $N=64$. To summarize,
this result indicates the minimum number of pixels that has to be
used experimentally to control molecular alignment with an optimal
control field.
\section{Conclusion}\label{sec4}
We have proposed a new monotonically convergent algorithm to take
into account spectral constraints. The procedure is built on the
standard framework but at each iteration, the field is taken as a
linear combination of the field given by the standard algorithm
and its filtered version. The parameter of the linear combination
is computed numerically to ensure the monotonic behavior of the
algorithm.

This algorithm has the advantage of simplicity and general
applicability whatever the filter chosen. We have presented two
examples on molecular alignment using either bandpass filters or a
discretization in the frequency domain mimicking pulse shaping
techniques. This work leads to important insights into the
different ways to achieve molecular alignment. Since there exists
no unique optimal solution, we can select using spectral filters
control fields taking into account experimental constraints. For
instance, we have shown that the control can find a pathway using
only three rotational frequencies to reach the aligned state. The
spectral constraints have suppressed the other pathways
simplifying therefore the structure of the control field.

This algorithm allows to test the choice of the filter with
respect to the optimal field considered. Indeed, a value of
$\mu_{k}$ close to 1 means that the filter is not well suited to
the form of the electric field determined by the monotonic
algorithm. This remark is particularly true for the first
iterations of the algorithm since after the first iterations, the
optimal field and the filtered field can be very close to each
other. In the case of pulse shaping techniques, we can then
determine from the algorithm the minimum number of pixels of the
mask needed to reach a high alignment efficiency. This renders the
use of optimal control theory more interesting from an
experimental point of view by making a direct link with control
experiments.

\appendix
\section{Proof of the monotonic behavior of the algorithm}\label{optapp}
In this section, we show the monotonic behavior of the algorithm
in the case of a nonlinear interaction with the control field. Let
$\tilde{E}_k$ be the optimal field at step $k$ of the algorithm.
We determine the field $E_{k+1}$ at step $k+1$ so that the
variation $\Delta J=J(E_{k+1})-J(\tilde{E}_k)\geq 0$
\cite{lapert}. This variation is given by:
\begin{equation}
\Delta J=|\langle \phi_f|\psi_{k+1}(t_f)\rangle|^2-|\langle
\phi_f|\psi_{k}(t_f)\rangle|^2-\int_0^{t_f}\lambda[E_{k+1}(t)^4-\tilde{E}_k(t)^4]dt.
\end{equation}
We introduce the function $P_{k+1}(t)=|\langle \chi_k(t)|\psi_{k+1}(t)\rangle|^2$. Differentiating it with respect to time produces
\begin{equation}
\frac{d}{dt}P_{k+1}(t)=\alpha_{k,k+1}(\tilde{E}_k^2-E_{k+1}^2)
\end{equation}
where
\begin{equation}
\alpha_{k,k+1}=2\textrm{Im}[\langle\psi_{k+1}(t)|\chi_k(t)\rangle\langle\chi_k(t)|\frac{1}{4}(\cos^2\theta\Delta \alpha+\alpha_\perp)|\psi_{k+1}(t)\rangle,\nonumber
\end{equation}
and a direct integration gives
\begin{equation}
P_{k+1}(t_f)=P_{k+1}(0)+\int_0^{t_f}\frac{dP_{k+1}}{dt} dt.
\end{equation}
Using the relation $|\langle \phi_f|\psi_{k+1}(t_f)\rangle|^2-|\langle \phi_f|\psi_{k}(t_f)\rangle|^2=P_{k+1}(t_f)-P_k(t_f)$, straightforward computations lead to
\begin{equation}
J(E_{k+1})-J(\tilde{E}_k)=-\int_0^{t_f}[\lambda(t)(E_{k+1}^4-\tilde{E}_k^4)-(\tilde{E}_k^2-E_{k+1}^2)\alpha_{k,k+1}]dt.
\end{equation}
The integrand $\mathcal{P}$ of $J(E_{k+1})-J(\tilde{E}_k)$ can be written as follows:
\begin{equation}
\mathcal{P}=(E_{k+1}-\tilde{E}_k)[-\lambda(E_{k+1}^3+E_{k+1}^2\tilde{E}_k+E_{k+1}\tilde{E}_k^2+\tilde{E}_k^3)-\alpha_{k,k+1}(E_{k+1}+\tilde{E}_k)]\nonumber.
\end{equation}
We define the optimal electric field $E_{k+1}$ as the solution of:
\begin{equation}
E_{k+1}-\tilde{E}_k=\eta [-\lambda(E_{k+1}^3+E_{k+1}^2\tilde{E}_k+E_{k+1}\tilde{E}_k^2+\tilde{E}_k^3)-\alpha_{k,k+1}(E_{k+1}+\tilde{E}_k)]\nonumber,
\end{equation}
where $\eta$ is a positive constant. It is straightforward to check that $J(E_{k+1}) - J(\tilde{E}_k) \geq 0$ for this choice of $E_{k+1}$.
\section*{Acknowledgements} D. S. acknowledges support from the Agence Nationale de la Recherche (ANR CoMoc). G.T. acknowledges financial support from C-QUID project and INRIA Rocquencourt.
%\bibliography{biblio}
%\bibliographystyle{apsrev}

\end{document}